\documentclass[12pt]{article}
\def\be{\begin{equation}}
\def\ee{\end{equation}}
\def\bea{\begin{eqnarray}}
\def\eea{\end{eqnarray}}
\usepackage{graphicx}

\catcode`\@=11
\def\lsim{\mathrel{\mathpalette\@versim<}}
\def\gsim{\mathrel{\mathpalette\@versim>}}
\def\@versim#1#2{\vcenter{\offinterlineskip
\ialign{$\m@th#1\hfil##\hfil$\crcr#2\crcr\sim\crcr } }}
\catcode`\@=12

\catcode`@=12
\begin{document}

\thispagestyle{empty}
\begin{flushright}
UCRHEP-T513\\
February 2012\
\end{flushright}
\vspace{0.3in}
\begin{center}
{\LARGE \bf Supersymmetric Axion-Neutrino Model\\
with a Higgs Hybrid\\}
\vspace{1.2in}
{\bf Ernest Ma\\}
\vspace{0.2in}
{\sl Department of Physics and Astronomy, University of California,\\ 
Riverside, California 92521, USA\\}
\end{center}
\vspace{1.2in}
\begin{abstract}\
In 2001, a supersymmetric model was propsed to relate the axion scale to 
that of neutrino mass seesaw.  Whereas this scenario is realistic, 
the particles associated with this mechanism are either too heavy or 
too weakly coupled for them to be observed (other than the axion itself 
or perhaps the axino).  A variation of that model is here proposed 
which allows significant mixing of the Higgs boson with a new singlet 
related to the saxion (the scalar partner of the pseudoscalar axion), 
rendering it possible to be observed at the Large Hadron Collider (LHC).  
With the addition of exotic color superfields, this also becomes a 
specific realization of how the production of such a 
Higgs hybrid may be suppressed or enhanced at the LHC, which is 
very relevant to ongoing experimental efforts to find the Higgs boson.
\end{abstract}

\newpage
\baselineskip 24pt

\section{Introduction}

Neutrino masses have been firmly established experimentally in recent 
years~\cite{mstv04}.  Any proposed model of particle interactions must now 
take this into account as a matter of course.  Yet there are two theoretical 
questions which are often not considered in connection with neutrino mass.  
One is the hierarchy problem and the other is the strong CP problem.  The 
first may be resolved by supersymmetry~\cite{m97} and the second by the 
spontaneous breaking of a Peccei-Quinn symmetry, i.e. $U(1)_{PQ}$, which 
results in a very light pseudoscalar boson, the axion~\cite{rvb00}.  Whereas 
the existence of supersymmetry is being explored at the Large Hadron Collider 
(LHC), and should be discovered if its breaking scale is around 1 TeV, the 
ongoing experimental search for the axion is yet to reach definitive limits. 

In 2001, a supersymmetric model was proposed~\cite{m01} to relate the axion 
scale to the anchor scale of the seesaw mechanism for neutrino mass.  The 
former is due to the vacuum expectation of a superfield in the range 
$10^9$ to $10^{12}$ GeV and the latter is the mass of a singlet neutrino 
superfield which could be somewhat smaller because it is multiplied by 
a Yukawa coupling.  The model is constructed with appropriate $U(1)_{PQ}$  
charges for three singlet superfields, so that the breaking of $U(1)_{PQ}$ 
at the very high axion scale preserves the supersymmetry which breaks at the 
much lower TeV scale.  This is then a comprehensive model of neutrino mass 
which is also intimately connected to the hierarchy and strong CP problems. 

The three singlet superfields $\hat{S}_{0,1,2}$ of this model are the new 
additions to the Minimal Supersymmetric Standard Model (MSSM) with also three 
singlet neutrino superfields $\hat{N}^c_{1,2,3}$.  The $U(1)_{PQ}$ symmetry 
is broken by $\langle S_{0,1} \rangle$, so that $N^c_{1,2,3}$ acquire large 
Majorana masses at the axion scale, but without breaking the supersymmetry. 
At the TeV scale of supersymmetry breaking, $\langle S_2 \rangle$ becomes 
nonzero and the $\mu$ term of the MSSM is generated.  The axion, axino, and 
saxion (the scalar partner of the axion) form the superfield which is 
a linear combination of $S_{0,1,2}$, whereas the other two linear combinations 
are very heavy.  Both the axino and saxion have masses at or 
below the scale of supersymmetry breaking, but are very weakly coupled 
to the usual particles of the MSSM.  As such, the LHC will not be very 
sensitive to the presence of these particles.  A possible exception is 
the case where the axino is the lightest particle of odd $R$ parity. 
In that case, it will contribute to the dark-matter relic density 
together with the axion, and the lightest MSSM neutralino will decay 
into it.  As for the saxion of this model, there is no practical way 
to find it at all.

In this paper, a new singlet superfield $\hat{S}_4$ is added and the 
$U(1)_{PQ}$ charges are redefined so that the $S_4$ scalar field now mixes 
significantly with the MSSM neutral scalars.  If $S_4$ also couples to heavy 
exotic color fermions, then this is a specific realization of the recently 
proposed idea~\cite{m11} that a Higgs hybrid may have a suppressed or 
enhanced production from gluon fusion at the LHC, which is very relevant 
for the ongoing experimental effort to find the Higgs boson, even if the 
125 GeV hint at the LHC is confirmed.  More detailed studies of this 
basic idea have recently appeared~\cite{bfh11, dkm11, eprzz11}.

In Sec.~2 the new model is defined.  In Sec.~3 the new mass spectrum and 
the new particles with masses at or below the supersymmetry breaking scale 
are discussed.  In Sec.~4 the possible new heavy exotic color particles 
are considered and the Higgs hybrid scenario is presented for either a 
suppression or enhancement of its production at the LHC.  In Sec.~5 there 
are some concluding remarks.

\section{Model}

The superfields of this model with their $U(1)_{PQ}$ charges are listed 
below.
\begin{table}[htb]
\begin{center}
\begin{tabular}{|c|c|c|}
\hline
superfield & $SU(3)_C \times SU(2)_L \times U(1)_Y$ & $U(1)_{PQ}$ \\ 
\hline
$\hat{Q}=(\hat{u}, \hat{d})$ & $(3,2,1/6)$ & $-1$ \\ 
$\hat{u}^c$ & $(3^*,1,-2/3)$ & $-1$ \\ 
$\hat{d}^c$ & $(3^*,1,1/3)$ & $-1$ \\ 
\hline
$\hat{L}=(\hat{\nu}, \hat{e})$ & $(1,2,-1/2)$ & $-3$ \\ 
$\hat{e}^c$ & $(1,1,1)$ & 1 \\ 
$\hat{N}^c$ & $(1,1,0)$ & 1 \\ 
\hline
$\hat{\phi}_1=(\hat{\phi}_1^0, \hat{\phi}_1^-)$ & $(1,2,-1/2)$ & 2 \\ 
$\hat{\phi}_2=(\hat{\phi}_2^+, \hat{\phi}_2^0)$ & $(1,2,1/2)$ & 2 \\ 
\hline
$\hat{S}_0$ & $(1,1,0)$ & $-2$ \\ 
$\hat{S}_1$ & $(1,1,0)$ & $-1$ \\ 
$\hat{S}_2$ & $(1,1,0)$ & 2 \\ 
$\hat{S}_4$ & $(1,1,0)$ & $-4$ \\ 
\hline
$\hat{\psi}$ & $(6,1,1/3)$ & 2 \\ 
$\hat{\psi}^c$ & $(6^*,1,-1/3)$ & 2 \\ 
\hline
\end{tabular}
\caption{$U(1)_{PQ}$ charges for all the superfields of this model.}
\end{center}
\end{table}
In this notation, all fields are left-handed.  The MSSM superfields are 
the usual $\hat{Q}, \hat{u}^c, \hat{d}^c, \hat{L}, \hat{e}^c, \hat{\phi}_1, 
\hat{\phi}_2$.  The new superfields are $\hat{N}^c$ and $\hat{S}_{0,1,2,4}$. 
The color sextets $\hat{\psi}, \hat{\psi}^c$ are to be discussed in 
Sec.~4.

As a result, the superpotential of this model is given by
\begin{eqnarray}
\hat{W} &=& m_2 \hat{S}_0 \hat{S}_2 + f_1 \hat{S}_1 \hat{S}_1 \hat{S_2} + 
f_2 \hat{S}_2 \hat{S_2} \hat{S}_4 + f_N \hat{S_0} \hat{N}^c \hat{N}^c 
+ f_4 \hat{S}_4 \hat{\phi}_1 \hat{\phi}_2 \nonumber \\ 
&+& h_d \hat{\phi}_1 \hat{Q} 
\hat{d}^c + h_u \hat{\phi}_2 \hat{Q} \hat{u}^c + h_e \hat{\phi}_1 \hat{L} 
\hat{e}^c + h_N \hat{\phi}_2 \hat{L} \hat{N}^c \nonumber \\ 
&+& h_1 \hat{\psi} \hat{u}^c \hat{d}^c + h_2 \hat{\psi}^c \hat{Q} \hat{Q} 
+ h_4 \hat{S}_4 \hat{\psi} \hat{\psi}^c.
\end{eqnarray}
Note that multiplicative lepton number $(-1)^L$ and baryon number $B$ are 
exactly conserved, with $\psi$ being a diquark superfield having $B=2/3$. 
The scalar potential coming from the first three terms of $\hat{W}$ is 
\begin{equation}
V = |m_2 S_2|^2 + |2f_1 S_1 S_2|^2 + |m_2 S_0 + f_1 S_1^2 + 2 f_2 S_2 S_4|^2 
+ |f_2 S_2^2|^2.
\end{equation}
Hence a solution which breaks $U(1)_{PQ}$ spontaneously while preserving 
supersymmetry is
\begin{equation}
\langle S_2 \rangle = \langle S_4 \rangle = 0, ~~~ m_2 \langle S_0 \rangle + 
f_1 \langle S_1 \rangle^2 = 0.
\end{equation}
As shown in Ref.~\cite{m01}, as the supersymmetry is broken at the $M_{SUSY}$ 
scale of about 1 TeV, this solution becomes
\begin{equation}
v_2 \sim v_4 \sim M_{SUSY}, ~~~ m_2 v_0 + f_1 v_1^2 \sim M_{SUSY}^2.
\end{equation}
Of the four superfields, two remain massive at the $m_2$ scale, the 
other two have masses of order $M_{SUSY}$, except for the very light axion. 
The axion scale is $\sqrt{4v_0^2 + v_1^2} \sim m_2$.  The heavy neutrino 
singlets have masses $f_N v_0$ and act as anchors in the usual Type I seesaw 
mechanism for very small Majorana masses of the active neutrinos. 
Variations of this basic idea is also possible for Type II, Type III, and 
radiative seesaw neutrino masses~\cite{m09}. 
The supersymmetric $\mu \hat{\phi}_1 \hat{\phi_2}$ term is generated with 
$\mu = f_4 v_4$, and the scalar $S_4$ mixes with the usual neutral Higgs 
scalars of the MSSM.  At the same time, the exotic color sextet quarks 
have masses $h_4 v_4$.  Whereas the MSSM Higgs scalars couple to two 
gluons through the SM quarks, the $S_4$ scalar couples to two gluons 
through the sextet quarks.

\section{New particles at the TeV Scale}

The axion comes from a linear superposition of the angular fields $\theta_i$ 
of $S_i = (1/\sqrt{2})(v_i + \rho_i) \exp(i \theta_i/v_i)$ as well as 
$\phi^0_{1,2} = (1/\sqrt{2})(v_{d,u} + \rho_{d,u}) \exp(i \theta_{d,u}/v_{d,u})$, 
i.e.
\begin{equation}
a = (-2 v_0 \theta_0 - v_1 \theta_1 + 2 v_2 \theta_2 - 4 v_4 \theta_4 
+ 2 v_d \theta_d + 2 v_u \theta_u)/V,
\end{equation}
where $V^2 = 4 v_0^2 + v_1^2 + 4 v_2^2 + 16 v_4^2 + 4 v_d^2 + 4 v_u^2$. 
In the presence of electroweak symmetry breaking, the combination 
$\theta_d \cos \beta - \theta_u \sin \beta$, where $\tan \beta = 
v_u/v_d$,  is absorbed by the $Z$ boson.  As a result, the electroweak 
component of the physical axion becomes $2 \sin 2 \beta (\theta_d 
\sin \beta + \theta_u \cos \beta)$.  The normalization $V$ is corrected by 
changing $(v_d^2 + v_u^2)$ to $(v_d^2 + v_u^2) \sin^2 2 \beta$. 
Since $v_{d,u} \sim 100$ GeV, $v_{2,4} \sim 1$ TeV, and $v_{0,1} \sim 10^9$ to 
$10^{12}$ GeV, the two massive superfields are roughly $\hat{S}_2$ and 
$(v_1 \hat{S}_0 - 2 v_0 \hat{S}_1)/V$, thereby allowing the superfield 
containing the axion and $\hat{S}_4$ to have components with masses of 
order $M_{SUSY}$.  Now this model is constructed with $\hat{S}_4$ 
interacting with $\hat{\phi}_1 \hat{\phi}_2$ and $\hat{\psi} \hat{\psi}^c$. 
The scalar $S_4$ will then mix with the scalar $\phi^0_{1,2}$, and 
so do their fermionic counterparts.  Such is of course a very 
familiar scenario in supersymmetry, where a singlet superfield is added 
to the MSSM.  Here it has an axion connection and may also couple to 
exotic quarks. 

The usual $4 \times 4$ mass matrix for the neutralinos of the MSSM is now 
extended to an $8 \times 8$ mass matrix, spanning $(\tilde{B}, \tilde{W}_3, 
\tilde{\phi}^0_1, \tilde{\phi}^0_2, \tilde{S}_0, \tilde{S}_1, \tilde{S}_2, 
\tilde{S}_4)$:
\begin{equation}
{\cal M} = \left[ \begin{array}
{c@{\quad}c@{\quad}c@{\quad}c@{\quad}c@{\quad}c@{\quad}c@{\quad}c}
\tilde{m}_1 & 0 & -sm_3 & sm_4 & 0 & 0 & 0 & 0 \\ 
0 & \tilde{m}_2 & cm_3 & -cm_4 & 0 & 0 & 0 & 0 \\ 
sm_3 & -cm_3 & 0 & f_4 v_4 & 0 & 0 & 0 & f_4 v_u \\ 
-sm_4 & cm_4 & f_4 v_4 & 0 & 0 & 0 & 0 & f_4 v_d \\ 
0 & 0 & 0 & 0 & 0 & 0 & m_2 & 0 \\ 
0 & 0 & 0 & 0 & 0 & 2 f_1 v_2 & 2 f_1 v_1 & 0 \\ 
0 & 0 & 0 & 0 & m_2 & 2 f_1 v_1 & 2 f_2 v_4 & 2 f_2 v_2 \\
0 & 0 & f_4 v_u & f_4 v_d & 0 & 0 & 2 f_2 v_2 & 0 
\end{array} \right],
\end{equation}
where $s = \sin \theta_W$, $c = \cos \theta_W$, $m_3 = M_Z \cos \beta$, 
$m_4 = M_Z \sin \beta$, with $\tan \beta = v_u/v_d$.  Using $m_2 \simeq 
f_1 v_1^2/v_0$, it is clear that $\tilde{S}_2$ combines with $(v_1 \tilde{S}_0 
- 2v_0 \tilde{S}_1)/V$ to form a Dirac fermion of mass $f_1 v_1 V/v_0$, and 
the axino is mostly $(2v_0 \tilde{S}_0 + v_1 \tilde{S}_1)/V$ with mass 
$2 f_1 v_2 (4 v_0^2/V^2)$.  The singlino $\tilde{S}_4$ mixes with the MSSM 
neutralinos.  If $v_4 \sim 1$ TeV, then its mass is approximately 
$-2 f_4 v_u v_d/v_4$ which is of order 10 GeV.  If $v_4 \sim 100$ GeV, 
then it mixes significantly with the MSSM neutralinos and all are of 
order 100 GeV.  In either case, it could be an important component of 
dark matter.  If the axino is the lightest particle of odd 
$R \equiv (-1)^{3B+L+2j}$, then the decay of the lightest neutralino 
into the axino plus a Higgs boson (real or virtual) is very much 
suppressed and may be visible at the LHC as a displaced vertex or not at all. 

As for the scalar sector, this scenario is very much like the NMSSM, 
where a singlet superfield is added~\cite{m10} to the MSSM.  However, 
there is no $\kappa \hat{S}_4 \hat{S}_4 \hat{S}_4$ term in the 
superpotential here.  There are two $3 \times 3$ mass-squared matrices, 
one for the real parts of $\phi^0_1,\phi^0_2,S_4$ and the other for the 
imaginary parts. The latter has one zero eigenvalue, corresponding to the 
would-be Goldstone boson for the longitudinal component of the $Z$ boson. 
In the NMSSM, there would be another zero eigenvalue (axion) because 
$\kappa =0$, but not so here because the axion is mostly not in $S_4$, 
as explicitly stated already in Eq.~(5).  Significant mixing between 
the neutral Higgs bosons of the MSSM with $S_4$ is possible, 
and the lightest particle of this sector could very well be a Higgs 
hybrid, with a suppressed (or enhanced) coupling to two gluons 
as recently proposed~\cite{m11}, thus affecting the ongoing search for  
the Higgs boson at the LHC.  Even if the 125 GeV hint at the LHC is 
confirmed, its production cross section times decay branching fraction 
into $\gamma \gamma$ for example may not be exactly as the SM predicts. 
If it is smaller or larger, then this scenario may be the answer 
especially if it is larger.

\section{Color sextets}

As Eq.~(1) shows, the color sextet superfields $\hat{\psi},\hat{\psi}^c$ 
of this model obtain their masses from $\langle S_4 \rangle = v_4$, just as 
$\hat{\phi}_{1,2}$.  They should thus appear at the TeV scale.  They also 
contribute to the $S_4-$gluon$-$gluon one-loop amplitude, in analogy to the 
usual $H-$gluon$-$gluon one-loop amplitude which is dominated by the $t$ 
quark.  The Higgs hybrid, i.e. $H' = H \cos \theta - S_4 \sin \theta$, may 
then have a suppressed (or enhanced) coupling to two gluons, as recently 
proposed~\cite{m11}.  [The color factor for sextets is 5/2 here, instead 
of 3 for the octets considered there.]  This will affect its experimental 
production cross section  at the LHC.  Furthermore, since the color sextets 
also have electric charge, the $H' \to \gamma \gamma$ branching fraction is 
changed as well.  At present, LEP and LHC data constrain the SM Higgs boson 
$H$ to be between 115 and 130 GeV in mass, and there is a hint that it 
may be 125 GeV.  If it is not discovered in this range, or discovered 
at a level, either above or below what is expected in the SM, this 
exotic model would be a possible explanation. Even if the SM Higgs boson 
is eventually confirmed, i.e. $\theta=0$, the scalar $S_4$ may still be 
discovered through its gluon-gluon coupling at a higher mass.

The presence of exotic particles in general has many phenomenological 
implications~\cite{mrs99}.  Here because of the $U(1)_{PQ}$ symmetry, 
both $(-1)^L$ and $B$ are conserved.  Hence proton decay as well as 
neutron-antineutron oscillations are forbidden.  Since $\psi$ couples 
to $u$ and $d$ quarks (and not $d$ and $d$ or $u$ and $u$ quarks), 
there are no tree-level contributions to $K^0-\bar{K}^0$, $B^0-\bar{B}^0$, 
$B_s^0 - \bar{B}_s^0$, or $D^0-\bar{D}^0$ mixing.  However, these mixings 
do occur in one-loop.  Whereas they should not contribute much to 
$K^0-\bar{K}^0$ mixing, they could have some nonnegligible effect 
on the others, especially $D^0-\bar{D}^0$~\cite{ghpp07}.  They also 
contribute to QCD penguin digrams and may be important for observing 
CP violation beyond the SM in charm decays for example.

\section{Concluding remarks}

To have a comprehensive model of particle interactions, a supersymmetric 
extension of the standard model is considered with an $U(1)_{PQ}$ symmetry, 
which is spontaneously broken (thus solving the strong CP problem) 
at a high scale by a set of singlet superfields without breaking the 
supersymmetry.  The neutrino singlet superfields also acquire mass at 
the same time, relating thus the seesaw mechanism for naturally small 
Majorana neutrino masses to the axion scale.  The dark matter of the 
Universe may have two components, the axion which has even $R$ parity, 
and the axino or neutralino which have odd $R$ parity.  All these were 
accomplished in a model proposed many years ago~\cite{m01}.  A new  
singlet superfield $\hat{S}_4$ is added here, which also couples to 
exotic color sextet diquarks, realizing thus the Higgs hybrid scenario 
proposed recently~\cite{m11} which allows for deviations from the 
standard-model predictions of Higgs boson production and decay. 
As LHC data accumulate, this scenario will be tested and its parameters 
constrained.

\section*{Acknowledgment}

This research is supported in part by the U.~S.~Department of Energy 
under Grant No.~DE-AC02-06CH11357.

\baselineskip 16pt
\bibliographystyle{unsrt}

\end{document}